\documentclass[prl,nofootinbib,twocolumn]{revtex4}

\usepackage{amsmath,epsfig}

\begin{document}

\title{Top Compositeness and Precision Unification}

\author{Kaustubh Agashe}
\author{Roberto Contino}
\author{Raman Sundrum}
\affiliation{Department of Physics and Astronomy, 
Johns Hopkins University, Baltimore, MD 21218-2686}


\begin{abstract}
The evolution of Standard Model gauge couplings is studied in a non-supersymmetric 
scenario in which the hierarchy problem is resolved by Higgs compositeness 
above the weak scale. It is argued that massiveness of the top quark combined 
with precision tests of the bottom quark imply that the right-handed 
top must also be composite. If, further, 
the Standard Model gauge symmetry is embedded 
into a simple subgroup of the unbroken composite-sector flavor symmetry, 
then precision coupling unification is shown to occur at~$\sim 10^{15}$ GeV,
to a degree comparable to supersymmetric unification.
\end{abstract}


\maketitle

The ambitious ideas of grand unification~\cite{Eidelman:2004wy}, 
and variants such as 
string unification~\cite{Dixon:1985jw} and orbifold unification~\cite{orbifoldGUT}, 
are founded on the structure and 
successes of the Standard Model~(SM). A central quantitative prediction is 
the evolution of SM gauge couplings from a single unified coupling, 
$\alpha_U$, at a unification scale, $M_U$. Schematically, in the minimal 
scenario, 
\begin{equation}
\label{SM}
\alpha_{i = 1, 2, 3}(\mu) = \alpha_U + \text{SM} +
 M_U\text{-physics}\, , 
\end{equation}
where the second term represents SM running, while the third 
represents model-dependent threshold effects from unification physics 
at a scale $\sim M_U$. Fortunately, it 
is natural for these $M_U$-scale effects to be 
much smaller than SM running, unless the $M_U$-sector is very large or 
 has large non-degeneracies. Neglecting these effects allows one to test 
unification with just SM data. 

There are several shortcomings: (i) The couplings do not meet very precisely. 
This does not falsify unification since $M_U$-effects may be unexpectedly large, 
but a more precise meeting without invoking such effects would have been 
much stronger circumstantial evidence for unification. Nevertheless, the 
results are intriguing. (ii) With the best fit, 
$M_U \sim 10^{14}$ GeV, so large that there is certainly no prospect of  
experimentally verifying any unified symmetry.
(iii) The requisite SM extrapolation to such high $M_U$ results in a 
severe gauge hierarchy problem. 
(iv) This $M_U$ is still low enough that exchange of massive states  
can  result in excessive proton decay. Such states can however be avoided in 
string or orbifold unifications~\cite{Dixon:1985jw,orbifoldGUT}. 

By comparison, unification in the context of weak scale supersymmetry (SUSY) 
is a striking success. 
The coupling evolution is given schematically by
\begin{equation}
\label{MSSM} 
\alpha_{i}(\mu) = \alpha_U + \text{SM} + \text{superpartners} + 
M_U\text{-physics}\, , 
\end{equation}
and again can be tested neglecting $M_U$-effects:
(i) Adding the superpartner-induced
 running yields a high precision meeting of 
couplings. 
The level of precision can be quantified by the postdiction
$\delta _3 \equiv (\alpha_3^{theory }-\alpha_3^{exp})/\alpha_3^{exp} \sim 10 \%$ at 
the scale $m_Z$.~\footnote{Often 
in the SUSY literature $\delta_3$ 
is evaluated at $M_U$ as being a few percent. This must be multiplied by 
$\alpha_3(m_Z)/\alpha_3(M_U) \sim 2.5$ in order to compare at 
$m_Z$, as we do.}
This size of $\delta_3$ can be naturally accounted for by threshold effects
from $M_U$-physics.
(ii) While the unification scale $M_U \sim 10^{16}$ GeV is still 
high, we rely less on directly seeing the unified gauge symmetry given the 
stronger circumstantial evidence. (iii) SUSY can solve the gauge hierarchy 
problem, so that two 
important issues are addressed simultaneously. (iv) $M_U$ is high enough to 
adequately suppress proton decay. 
String unification or orbifold unification 
are still attractive for solving 
the doublet-triplet splitting problem~\cite{Dixon:1985jw, orbifoldGUT}.

In this letter we pursue a very different scenario, namely that the 
hierarchy problem is solved by having the Higgs doublet be a 
composite of some new (non-supersymmetric) strong 
dynamics~\cite{Kaplan:1983fs}.~\footnote{We 
will not follow the technicolor approach~\cite{Weinberg:1975gm}, in which a 
Higgs scalar is effectively absent.} 
While such dynamics is necessarily non-perturbative and theoretically 
challenging, there has been a 
recent revival of interest  because of two 
extensions which allow one to understand weak scale 
symmetry breaking, precision tests and phenomenology, {\it independent of  
many of the details of the strong sector}. Little Higgs theory 
\cite{Arkani-Hamed:2002qx} is one 
such extension, which we will  
not pursue here. Our work is motivated by 
(but not strongly reliant on) Refs. \cite{Agashe:2003zs, Agashe:2004rs}, a realistic 
Randall-Sundrum (RS) 
extra-dimensional scenario with most or all of the SM fields
in the bulk. Via the AdS/CFT correspondence \cite{Aharony:1999ti,RSCFT}, 
such a scenario is 
{\it dual} to a purely 4D composite Higgs scenario. The Kaluza-Klein 
excitations map to some low-lying hadrons  at 
the compositeness scale, $\Lambda_{comp}$. 
 The ratio 
of higher-dimensional curvature to the effective field theory cutoff 
maps to a new small parameter of the strong sector, with the help of 
which many weak-scale 
observables can be calculated independently of microscopic details 
of the strong dynamics. 

An attractive feature of (the 4D dual of) this type of RS
 set-up is its simple extrapolation (at least for some important 
inclusive observables) 
to energies far above the weak scale. This leads to an elegant mechanism for 
generating hierarchical Yukawa couplings~\cite{Kaplan:1991dc,Grossman:1999ra,%
Gherghetta:2000qt, Huber:2000ie}. 
The light SM fermions are taken to 
be elementary particles, weakly coupled  
 to strong-sector operators. Running down to
$\Lambda_{comp}$, these operators
 induce small Yukawa couplings to the Higgs composite. 
Hierarchies arise 
naturally from the different scaling dimensions
 of different strong operators. The weak couplings  to 
the strong sector also naturally suppress modifications of couplings to the 
$W, Z$ \cite{Huber:2000fh} and 
compositeness and flavor-changing effects in the light SM 
fermions \cite{Gherghetta:2000qt}, in accord 
with modern data. 

The top quark is, however, a special case. Its Yukawa coupling to the Higgs 
composite is so large that either $t_R$, $t_L$, or both must effectively 
also be composite. However, precision data such as tests of 
$Z \rightarrow b \overline{b}$  strongly suggest that $b_L$, and hence 
$t_L$ by 
electroweak symmetry, can have at most a small admixture of a TeV-scale 
composite~\cite{Georgi:1994ha}\cite{Agashe:2003zs}. 
We deduce that $t_R$ must be the composite. 

With these broad motivations and expectations, 
we will show that under quite simple 
and plausible conditions an attractive scheme for precision 
unification emerges. We will first derive our central result to
leading order (LO) in the couplings of elementary fields to the strong sector,  
and then consider  subdominant corrections. 
Our discussion will be mostly from the 4D 
viewpoint of the strong sector, rather than an RS description, 
for reasons we will explain later. Closely related ideas in a SUSY RS 
context were presented in Ref.~\cite{Gherghetta:2004sq}.~\footnote{Other 
RS unification proposals not tied to top/Higgs compositeness appear in 
Refs.~\cite{Pomarol:2000hp, Randall:2001gb, Goldberger:2002pc}.}

Since 
composites carrying electroweak quantum numbers and color
emerge from the new strong sector, the SM vector fields 
must gauge some $SU(3) \times SU(2) \times U(1)$ subgroup of 
the global ``flavor'' group, $G$. Above $\Lambda_{comp}$, the running due to
the light composites, $H, t_R$, must be replaced by the full strong 
dynamics,  
\begin{equation}
\label{1top} 
\alpha_{i}(\mu) = \alpha_U + \text{SM} - \{t_R, H\} +
{\rm strong~sector} + M_U\text{-physics}. 
\end{equation}
Fortunately, the non-perturbative strong 
sector  contributions to SM running cancel to one-SM-loop order 
in computing {\it differential running}, 
that is the running of $(\alpha_i - \alpha_1)$ say, if the 
SM gauge group is embedded in a simple factor 
of G~\footnote{This also includes the case in which $G$ contains
some discrete generators. For example, $SU(N)_L\times SU(N)_R$
\textit{with} parity is simple.} (such as $SU(5)$ for example):
\begin{equation}
\label{1topdiff} 
\alpha_{i}(\mu) - \alpha_1(\mu) = \text{SM} - \{t_R, H\} +
M_U\text{-physics}\, .
\end{equation}
This is  all we need to check gauge coupling unification.  
We assume the simple embedding of the SM into $G$ from now on.

Eq.~(\ref{1topdiff}) exhibits a remarkable twist in the unification
paradigm. Instead of adding the running from physics beyond the SM, 
here compositeness instructs us to {\it subtract} the running due to some SM 
particles! Before checking unification we must 
explain why the light composites, $H, t_R$, do not fill out complete 
representations of the global symmetry, $G$.  There are two distinct cases
 following from the possibility of spontaneous symmetry 
breaking in the strong sector at $\Lambda_{comp}$, $G \rightarrow K$.
(Indeed the original, and still attractive, proposal for a composite Higgs
is as a (pseudo-)Goldstone boson of this type of symmetry breaking 
\cite{Kaplan:1983fs}, though 
this is not essential for the present paper.) The two 
cases are (a) the SM gauge group remains embedded in a 
simple factor of $K \subset G$, or (b) it does not.
In (b) there is no contradiction 
with $H, t_R$ being the only light composites, and 
Eq.~(\ref{1topdiff}) applies. One finds
that the 
subtractions certainly 
improve unification, but it is still 
not very precise. We will not study this case further here.

Here, we focus on (a), where $H, t_R$ {\it must} be accompanied 
by other composites, filling out complete $K$-representations. Having extra 
(colored) scalar composites does not pose a robust problem, since  
the perturbations of the SM coupling to the strong sector can easily split 
the Higgs doublet from its $K$-partners, allowing the former to condense and 
be light while the latter do not condense and are massive enough to avoid 
present bounds. But the chiral fermionic 
$K$-partners of the $t_R$ do pose a robust problem (and introduce SM 
anomalies). The only way to remove 
these unwanted states is to assume there exist exotic elementary fermions 
beyond the SM with couplings to the strong sector, which induce Dirac masses 
with the $K$-partners of the $t_R$ below $\Lambda_{comp}$. That is, the 
exotics must have SM quantum numbers which are charge-conjugate to the 
$K$-partners of the $t_R$ \cite{Agashe:2003zs, Contino:2004vy}. 

\begin{figure}[t]
\centering
\epsfig{file=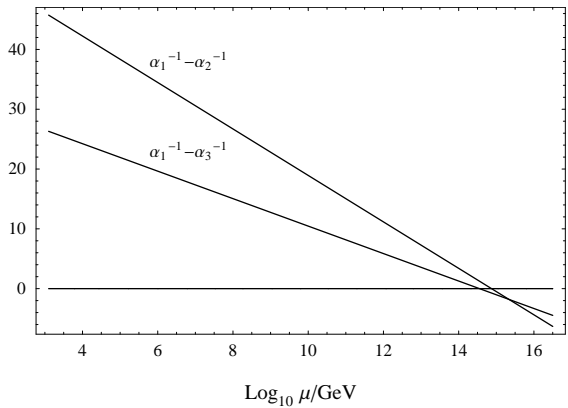,width=\linewidth}
\caption{\label{fig:unifLO} LO differential running of SM gauge couplings 
in the top/Higgs compositeness scenario (a).}
\end{figure}

The elementary exotics also contribute to SM running
above $\Lambda_{comp}$,
\begin{equation}
\begin{split}
\label{2top} 
\alpha_{i}(\mu) = &\alpha_U + \text{SM} - \{t_R, H\} + 
\text{exotics} \\
 &+ \text{strong sector} + M_U\text{-physics}\, . 
\end{split}
\end{equation}
We assume here that 
 the exotic couplings to the strong sector 
are  weak enough that their contributions to SM running are
approximately undressed by 
strong-sector corrections.  At one-SM-loop the 
{\it differential} running only depends on the fact that the exotics fill out 
a complete $K$-representation {\it except for a missing} $t_R^c$, 
\begin{equation}
\label{2topdiff} 
\alpha_{i}(\mu) - \alpha_1(\mu) = \text{SM} - \{t_R, t_R^c, H\} +
M_U\text{-physics}\, . 
\end{equation}
Neglecting the $M_U$-threshold, as usual, yields near perfect unification,  
with $M_U \sim 10^{15}$ GeV. See Fig.~\ref{fig:unifLO}.

Eq.~(\ref{2topdiff}) and Fig.~\ref{fig:unifLO} summarize our central quantitative result
to LO in SM gauge 
couplings and zeroth order in the 
couplings of elementary fermions to the strong sector.
We now discuss the subleading corrections. It is difficult to couple 
elementary fermions to strong sector operators at $M_U \sim 10^{15}$ GeV
without the couplings being highly irrelevant in the IR, resulting in 
negligible Yukawa couplings, 
 unless {\it the strong 
sector is  strongly-coupled throughout the large hierarchy}. This happens 
naturally when the  
strong sector is near an IR-attractive fixed point above  
$\Lambda_{comp}$. In this case, 
working to next-to-leading order (NLO),
the gauge coupling running above $\Lambda_{comp}$ 
is given by 
\begin{equation}
\label{RG}
\frac{d}{d \ln \mu} \left(\frac{1}{\alpha_i}\right) = \frac{b_i}{2 \pi} 
+ \frac{B_{ij}}{2 \pi} \frac{\alpha_j}{4 \pi} +  \frac{C_{i \alpha}}{2 \pi} 
\frac{\lambda_{\alpha}^2}{16 \pi^2}\, ,
\end{equation}
where the $b, B, C$ are constants
and $\lambda_{\alpha}, \alpha = exotic, Q_L^3 \equiv (t_L, b_L)$, 
denote the largest couplings of the 
elementary fermions to the strong 
sector (resulting in the largest masses with composite fermions). 
We further decompose
\begin{equation}
\begin{split}
b_i &\equiv b_i^{SM-} + b_i^{exotic} + b^{strong}  \\
B_{ij} &\equiv B_{ij}^{SM-} + B_{ij}^{exotic} + B_{ij}^{strong}\, , 
\end{split}
\end{equation}
where ``$SM-$'' refers to $SM - \{t_R, H\}$.
Note that  $b_i^{SM-},  b_i^{exotic}, B_{ij}^{SM-},  B_{ij}^{exotic}$ are 
just representation-theoretic factors. For concreteness we consider 
 the SM gauge 
group embedded in $SO(10) \subset K$ in the usual way, with the $t_R$ 
being part of a composite $16$ of $SO(10)$, so that the elementary exotics  
$\equiv \overline{16} - \{t_R^c\}$. 

By contrast, 
 $b^{strong}, B_{ij}^{strong}, 
C_{i \alpha}$ include unknown ${\cal O}(1)$ strong interaction factors. 
We will treat $b^{strong}$ as an unknown ($i$-independent by 
the SM embedding into a simple factor of $G$),
which can usefully be thought of as a crude measure of the $SO(10)$-charged
content of the strong sector.
A rough but reasonable expectation is that $b^{strong}\gtrsim b^{comp}$,
where $b^{comp}$ is the LO renormalization group coefficient due to 
the light composites alone in the far IR.
For a real scalar 10 and a Weyl fermion 16 of $SO(10)$ ($\ni H, t_R$), 
$b^{comp}=1.5$.
We will use crude estimates of  
the NLO coefficients, $B_{ij}^{strong} \sim 3 \cdot 3 \cdot b^{strong}$, 
$C_{i \alpha} \sim 3 \cdot b^{strong}$, 
as part of our theoretical error 
estimates. These estimates follow from the fact that the (non-perturbative) 
diagrams contributing to 
the NLO coefficients arise from diagrams contributing to 
$b^{strong}$ with insertions of intermediate elementary gauge bosons or 
fermions via SM gauge couplings or $\lambda_{\alpha}$. Such insertions can 
result in summation over QCD colors or weak isospins, giving rise to an extra 
factor of at most $3$. Further, experience with perturbative gauge loops shows
that they give an extra factor $\sim 3$ beyond the naive loop-counting 
parameter. This accounts for the second factor of $3$ in $B^{strong}$. 

Following (the AdS/CFT dual of) the scenario of Ref.~\cite{Agashe:2003zs},
we assume that above 
$\Lambda_{comp}$ the 
couplings $\lambda_{\alpha}$ are 
slightly relevant, driving the theory
away from the original fixed point (of the isolated strong sector) to a 
nearby fixed point. We can approximate $\lambda_{\alpha}(\mu)$ in the 
gauge coupling running by their new-fixed-point values, 
$\lambda_{*\alpha}$. Integrating Eq.~(\ref{RG}) down to $\mu \lesssim
\Lambda_{comp}, m_{exotic}$, 
\begin{equation}
\label{solution}
\begin{split}
\frac{1}{\alpha_i(\mu )} =&
 \frac{1}{\alpha_U} - \frac{b_i}{2 \pi} \ln\frac{M_U}{\mu}
    - \frac{C_{i \alpha}}{32 \pi^3} \lambda_{*\alpha}^2 \ln\frac{M_U}{\mu} \\
  & - \frac{B_{ij}}{4 \pi b_j} \ln\left[1 + \alpha_j(\mu) \frac{b_j}{2 \pi} 
    \ln\frac{M_U}{\mu} \right] \\
  & +  \text{threshold corrections}\, .
\end{split}
\end{equation}

Let us first discuss the thresholds at $M_U \sim 10^{15}$~GeV, 
$\Lambda_{comp} \sim$ few TeV, and $m_{exotic} \sim$~TeV. 
The detailed physics at $M_U$ 
is unknown, but  the threshold effects 
associated with the strong sector can be
subsumed into $\alpha_U$, $B^{strong}$, $C$. 
As in standard unification schemes, the minimal natural size
of threshold effects associated with elementary fields
is $\delta(1/\alpha_i)\sim {\cal O}(1)/2\pi$.
The expectation $\Lambda_{comp} \sim$ few TeV follows from the requirements of
reasonable naturalness of the weak scale as well as passing  
electroweak precision tests. This was demonstrated 
in RS modelling using the  
extra-dimensional calculability \cite{Agashe:2003zs,Agashe:2004rs}. 
In the present scenario, we do not expect a 
useful extra-dimensional dual description, as we explain later, 
but the RS
calculations of precision observable 
serve as plausible estimates, with at most 
${\cal O}(1)$ unknown correction factors. A central requirement however, is
having an approximate ``custodial isospin'' symmetry of the strong sector to 
 protect the electroweak $\rho$-parameter. Our choice of
$SO(10)$ flavor symmetry 
ensures this, with custodial $SU(2)_R$ as well 
as SM subgroups. As a consequence, one exotic has the gauge quantum numbers 
of $\nu_R$, is stable given a baryon number symmetry of the strong sector, 
and can serve as a dark matter candidate if its mass is 
$\sim$ few hundred~GeV~\cite{Agashe:2004ci}. 
However, the exotic $SU(2)_R$-partner of $t_R$ must 
have mass $\gtrsim 1.5$ TeV in order to avoid it forming too large a component of
 the 
observed bottom quark, in contradiction with precision tests.
These considerations motivate $m_{exotic} \sim$ TeV, with mild 
$SO(10)$-violating splittings. A simple (but not the only) 
way for this to happen is for $SU(5)$ to be an exact flavor symmetry of the
strong sector, with the remainder of the
 $SO(10)$ symmetry being only approximate. Exact $SU(5)$ is all we need here.

In SUSY unification the one-loop superpartner threshold effects are generally 
significant because of large non-gauge-universal 
splittings in their spectrum induced by running from $M_U$,
and also
by the need to avoid existing search constraints and extreme fine-tuning. 
In the present scenario this does not happen because 
the $\Lambda_{comp}$ threshold approximately has the global $K$-symmetry of 
the strong sector, while the exotics and their Dirac partners also come 
in an almost $K$-symmetric form, with only the $t_R^c$ state missing. 
We will therefore probe our sensitivity to the associated
 threshold corrections by simply varying $\Lambda_{comp}$ 
from~$3 - 5$ TeV, and insert a single (for simplicity) 
exotic threshold from~$0.5 - 2$ TeV in running SM couplings measured 
at $m_Z$ up to $\Lambda_{comp}$.

Finally, we need to estimate the weakly-perturbed fixed-point couplings, 
$\lambda_{* \alpha}$. $\lambda_{* Q_L^3}$ is responsible for coupling the 
elementary $t_L$ to the strong sector, yielding a Yukawa coupling below 
$\Lambda_{comp}$ of $\lambda_{* Q_L^3}$ times an ${\cal O}(1)$ strong 
interaction factor. Thus we have $\lambda_{* Q_L^3} \sim 1$. 
 $\lambda_{* exotic}$
 is responsible for generating a Dirac mass for the exotics
with the excess fermion composites, $m_{exotic} \sim 
 \lambda_{* exotic} \Lambda_{comp} \sqrt{b^{strong}}/4 \pi $. For 
$m_{exotic} \sim {\cal }({\rm TeV})$ we also need $\lambda_{* exotic} \sim 1$.
Thus, in our analysis $\lambda_{* \alpha} \sim 1$. 

\begin{figure}[t]
\hspace*{-0.4cm}
\epsfig{file=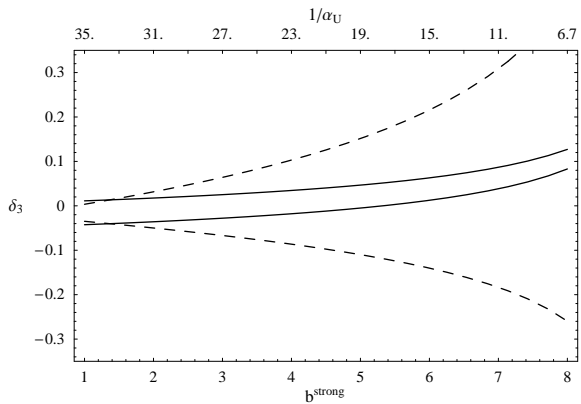,width=\linewidth}
\caption{\label{fig:unifNLO} 
Postdiction of $\alpha_3(m_Z)$ at NLO as 
given in terms of $\delta _3 = (\alpha_3^{theory }-\alpha_3^{exp})/\alpha_3^{exp}$.
Solid lines denote the error band from threshold effects at $M_U$,
$\Lambda_{comp}$, $m_{exotic}$, whereas dashed lines denote the estimated error from
$B_{ij}^{strong}$, $C_{i \alpha}$.} 
\end{figure}

In Fig.~\ref{fig:unifNLO} we exhibit a simple and
 standard test of unification, given the high precision 
of electroweak data, namely using measured values of $\alpha_{1, ~2}$ to 
postdict $\alpha_3(m_Z)$. We use separate bands to denote the variation 
in postdicted $\alpha_3(m_Z)$ coming from the above threshold ranges and 
from the theoretical error arising from our
$B_{ij}^{strong}, C_{i \alpha}$ bounds. 
Note that our central predictions are excellent, we do not {\it need} 
large corrections, our largest uncertainties 
just reflect the conservative bounds put on the $B^{strong}, C$.
The regime of controlled unification 
involves modest $b^{strong}$, not much larger than the size of the 
strong flavor group.  
Also, requiring that the SM gauge couplings do not have a Landau pole
below $M_U$ implies $b^{strong}\lesssim 9$.
This suggests that an AdS dual description, 
requiring a large ratio of strong colors ($\sim {\cal O}(b^{strong})$)
to flavors, will not be useful. This
is why we have not pursued RS modelling in the present context.

Let us assess our scenario with the criteria used to discuss 
earlier unification scenarios: (i) The couplings meet very precisely, 
strong circumstantial evidence 
for this form of unification. The postdiction of 
$\alpha_3(m_Z)$ works to better than $\delta_3 \sim 15\%$
over a wide range of $b^{strong}$. Alternatively, postdiction of 
sin$^2 \theta_W$ using $\alpha_3^{exp}$ leads to an error
$\delta \sin^2\theta_W \simeq  0.03\, \delta_3$.
This is quite comparable with the level of success of 
SUSY unification~\cite{Langacker:1992rq}. 
(ii) With the best fit, $M_U \sim 10^{15}$ GeV, so large that
experiments will not directly see the unification physics. 
However, there will be a striking signature of unification
surviving to accessible energies~\cite{Pomarol:2000hp},
namely the strong sector resonances will fill out approximately 
degenerate multiplets of a {\it unified} flavor symmetry, $K$.
Since their masses are expected to be in the
few TeV range (based on naturalness), the cross-section for their single
production at the LHC could be significant. These resonances will decay mostly into
$t_R$ and Higgs (including longitudinal $W/Z$) due to the strong coupling
involved, quite distinctly from
other models such as SUSY.
(iii) We have arrived at precision unification here by considering one of the 
simplest non-supersymmetric 
scenarios for solving the hierarchy problem of the SM. 
(iv) $M_U \sim 10^{15}$ GeV is still low enough that 
exchanges of $X,Y$ bosons at this scale
can result in excessive proton decay. 
Even more importantly, composite states with the same quantum numbers
can also mediate proton decay. 
The only known schemes in which both problems are solved are string
or orbifold unification~\cite{Dixon:1985jw, orbifoldGUT, Goldberger:2002pc}, so this is 
a requirement of our scenario. 
There could also be UV model-dependent states at $M_U$ contributing to proton decay.
These effects can be suppressed by imposing a (gauged) baryon-number symmetry, compatible with
orbifold unification, as long as it is broken somewhat below $M_U$. This symmetry
should also be an accidental flavor symmetry of the strong dynamics,
to extend the usual accidental baryon-number symmetry of the SM.
A second reason for preferring string/orbifold unification  
is that it makes it simpler to understand the appearance of incomplete 
grand-unified fermion multiplets in the IR, such as our exotics. 
In orbifold unification the 
global strong-sector symmetry $G$ (or $K$) may even be  
the grand unified gauge group, surviving orbifold projections in this sector, 
but not in the elementary fermion/gauge-boson sectors.
(v) As mentioned earlier, an attractive dark matter candidate emerges as 
a $K$-partner of $t_R$~\cite{Agashe:2004ci}.

\vspace*{0.3cm}
The scenario in which the SM hierarchy problem is solved 
non-supersymmetrically with top/Higgs compositeness, or a  RS
 dual depiction, 
is attractive from several phenomenological 
points of view. In this letter, we have studied one of the key features 
that has been taken as strong evidence in favor of a supersymmetric 
solution to the hierarchy problem, namely precision gauge-coupling 
unification. We have found an equally striking (but very different) 
unification that 
follows rather minimally from top/Higgs compositeness. 
We hope to have shown that  taking 
unification as a serious consideration, one must still 
keep an open mind as to 
how the hierarchy problem is resolved in Nature, supersymmetrically or 
non-supersymmetrically. This is not a passive state, 
extracting new physics from upcoming colliders is challenging and 
requires planning ahead.

\begin{acknowledgments}
We thank Tony Gherghetta and Alex Pomarol for discussions.
This work has been supported by NSF Grant P420-D36-2051.
\end{acknowledgments}


\begin{thebibliography}{99}


\bibitem{Eidelman:2004wy}
For a review, see 
S.~Raby in S.~Eidelman {\it et al.}  [Particle Data Group],
Phys.\ Lett.\ B {\bf 592}, 1 (2004).


\bibitem{Dixon:1985jw}
L.~J.~Dixon, J.~A.~Harvey, C.~Vafa and E.~Witten,

Nucl.\ Phys.\ B {\bf 261}, 678 (1985);
{\bf 274}, 285 (1986).


\bibitem{orbifoldGUT}
See
Y.~Kawamura,
Prog.\ Theor.\ Phys.\  {\bf 103}, 613 (2000); 
{\bf 105}, 999 (2001);
L.~J.~Hall and Y.~Nomura,
Annals Phys.\  {\bf 306}, 132 (2003)
and references therein.

\bibitem{Kaplan:1983fs}
D.~B.~Kaplan and H.~Georgi,
Phys.\ Lett.\ B {\bf 136}, 183 (1984).


\bibitem{Weinberg:1975gm}
S.~Weinberg,
Phys.\ Rev.\ D {\bf 13}, 974 (1976);
{\bf 19}, 1277 (1979);
L.~Susskind,
Phys.\ Rev.\ D {\bf 20}, 2619 (1979).


\bibitem{Arkani-Hamed:2002qx}
N.~Arkani-Hamed \textit{et al.},
JHEP {\bf 0208}, 021 (2002);
N.~Arkani-Hamed, A.~G.~Cohen, E.~Katz and A.~E.~Nelson,
JHEP {\bf 0207}, 034 (2002).
For a UV completion of the Littlest Higgs
model using AdS$_5$ see
J.~Thaler and I.~Yavin,
arXiv:hep-ph/0501036.




\bibitem{Agashe:2003zs}
K.~Agashe, A.~Delgado, M.~J.~May and R.~Sundrum,
JHEP {\bf 0308}, 050 (2003).


\bibitem{Agashe:2004rs}
K.~Agashe, R.~Contino and A.~Pomarol,
arXiv:hep-ph/0412089.


\bibitem{Aharony:1999ti}
For a review, see
O.~Aharony \textit{et al.},
Phys.\ Rept.\  {\bf 323}, 183 (2000).



\bibitem{RSCFT}
N.~Arkani-Hamed, M.~Porrati and L.~Randall,
JHEP {\bf 0108}, 017 (2001);
R.~Rattazzi and A.~Zaffaroni,
JHEP {\bf 0104}, 021 (2001).


\bibitem{Kaplan:1991dc}
D.~B.~Kaplan,
Nucl.\ Phys.\ B {\bf 365}, 259 (1991).

\bibitem{Grossman:1999ra}
Y.~Grossman and M.~Neubert,
Phys.\ Lett.\ B {\bf 474}, 361 (2000).

\bibitem{Gherghetta:2000qt}
T.~Gherghetta and A.~Pomarol,
Nucl.\ Phys.\ B {\bf 586}, 141 (2000).

\bibitem{Huber:2000ie}
S.~J.~Huber and Q.~Shafi,
Phys.\ Lett.\ B {\bf 498}, 256 (2001);
S.~J.~Huber,
Nucl.\ Phys.\ B {\bf 666}, 269 (2003).


\bibitem{Huber:2000fh}
S.~J.~Huber and Q.~Shafi,
Phys.\ Rev.\ D {\bf 63}, 045010 (2001).

\bibitem{Georgi:1994ha}
H.~Georgi, L.~Kaplan, D.~Morin and A.~Schenk,
Phys.\ Rev.\ D {\bf 51}, 3888 (1995).

\bibitem{Gherghetta:2004sq}
T.~Gherghetta,
arXiv:hep-ph/0411090.


\bibitem{Pomarol:2000hp}
A.~Pomarol,
Phys.\ Rev.\ Lett.\  {\bf 85}, 4004 (2000).

\bibitem{Randall:2001gb}
L.~Randall and M.~D.~Schwartz,
JHEP {\bf 0111}, 003 (2001);
K.~Agashe, A.~Delgado and R.~Sundrum,
Annals Phys.\  {\bf 304}, 145 (2003).

\bibitem{Goldberger:2002pc}
W.~D.~Goldberger, Y.~Nomura and D.~R.~Smith,
Phys.\ Rev.\ D {\bf 67}, 075021 (2003).

\bibitem{Contino:2004vy}
R.~Contino and A.~Pomarol,
JHEP {\bf 0411}, 058 (2004).



\bibitem{Agashe:2004ci}
K.~Agashe and G.~Servant,
Phys.\ Rev.\ Lett.\  {\bf 93}, 231805 (2004);
arXiv:hep-ph/0411254.

\bibitem{Langacker:1992rq}
For a NLO analysis, see
P.~Langacker and N.~Polonsky,
Phys.\ Rev.\ D {\bf 47}, 4028 (1993).




\end{thebibliography}
\end{document}